\documentclass[aps,superscriptaddress,twocolumn,a4paper,pra,showpacs]{revtex4-1}
\usepackage{latexsym,graphicx,amsmath,amsfonts}
\def\ket#1{| #1 \rangle}
\def\bra#1{\langle #1 |}
\def\proj#1{\ket{#1}\bra{#1}}
\def\tr{\text{Tr}}

\newcommand{\one}{\mathbb{I}}

\begin{document}
\title{Unlearning Quantum Information}

\author{Daniel~K.~L.~\surname{Oi}} \email{daniel.oi@strath.ac.uk}
\affiliation{SUPA Department of Physics, University of Strathclyde,
  Glasgow G4 0NG, United Kingdom}

\begin{abstract}
  Quantum dynamics can be driven by measurement. By constructing
  measurements that gain no information, effective unitary evolution
  can be induced on a quantum system, for example in ancilla driven
  quantum computation. In the non-ideal case where a measurement does
  reveal some information about the system, it may be possible to
  ``unlearn'' this information and restore unitary evolution through
  subsequent measurements. Here we analyse two methods of quantum
  ``unlearning'' and present a simplified proof of the bound on the
  probability of successfully applying the required correction
  operators. We find that the probability of successful recovery is
  inversely related to the ability of the initial measurement to
  exclude the possibility of a state.
\end{abstract}

\date{\today}


\maketitle

\section{Introduction}

In quantum information processing schemes such as measurement-based
quantum computation~\cite{MBQC2003}, ancilla-driven quantum
computation~\cite{Twisted2009,ADQC2010,Twisted2012}, and holonomic
degenerate projections~\cite{AP1989,Oi2014}, unitary quantum dynamics
are driven by measurements that learn nothing about the
system. Previous work has studied several issues including non-ideal
coupling between system and ancilla~\cite{MK2010}, preparation, gate,
storage and measurement errors~\cite{RHG2007}. Here, we address the
issue of ``unlearning'' information gained from a non-ideal
generalized measurement by subsequent conditional measurements to
restore the unitary evolution of the system. This question has been
addressed before in the context of reversing
measurement~\cite{UIN1996,KU1999,ParaoanuPRA2011,CL2012}, experimental
proposals~\cite{KJ2006,JK2010,ParaoanuEPL2011} and
demonstrations~\cite{Katz2008,Kim2009}. We present a simplified proof
of the bound on the success probability of such corrective measures
and relate it to the spectrum of the measurement operators
corresponding to non-unitary evolution as well as describing finite
and asymptotic correction schemes achieving this limit.

\section{Preliminaries}

A generalized measurement, or positive operator valued measure (POVM),
can be described by a set of positive operators $\{M_j\}$ that sum to
the identity, $\sum_j M_j=\one$. The probability of obtaining outcome
$j$ when measuring system described by density operator $\rho$ is
$p_j=\tr[M_j \rho]$. The post-measurement state is not uniquely
defined by $M_j$ in general, but is given by $\rho_j=\frac{K_j\rho
  K_j^\dagger}{\tr[K_j^\dagger K_j \rho]}$ where $M_j=K_j^\dagger
K_j$, and $\{K_j\}$ are Kraus operators. In this paper, we will always
consider the post-measurement state to be of the same dimensionality
as the input.

\begin{figure}
\includegraphics[width=\columnwidth]{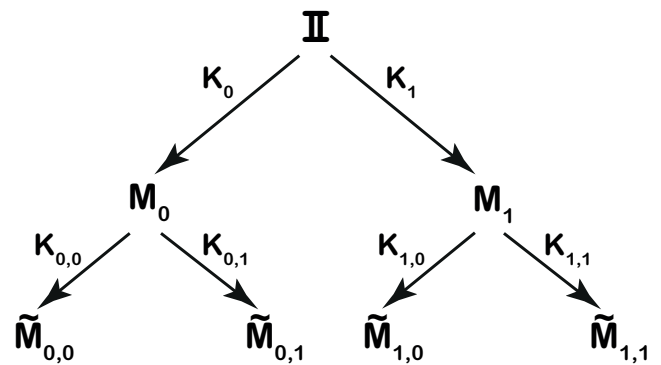}
\caption{Two-level binary POVM tree. Each bifurcation represents a
  binary POVM with two Kraus operators labeling the arrows. The nodes
  of the tree represent the cumulative measurement operator
  corresponding to the sequence of results leading to that node. The
  cumulative Kraus operator consists of the product of all the Kraus
  operators along the path down the branch. The sum of the $\tilde{M}$
  children of a branch sum up to parent node.}
\label{fig:povmtree}
\end{figure}

A cascaded sequence of measurements (Fig.~\ref{fig:povmtree}) results
in a cumulative Kraus operator that is the product of the individual
Kraus operators associated with each sequential result, e.g. if a
first measurement has Kraus operators $\{K_j\}$, and depending on the
result $j$, a second measurement is performed with Kraus operators
$\{K_{j,k}\}$, the total Kraus operator associated with joint result
$j$ and then $k$ is given by $\tilde{K}_{j,k}=K_{j,k}K_{j}$, and the
POVM element is given by $\tilde{M}_{j,k}=(\tilde{K}_{j,k})^\dagger
\tilde{K}_{j,k}$.

We will consider the case where ideally we would like the Kraus
operators to be proportional to a unitary, $K_j=q_j U_j$ where
$0<q_j\le 1$ for some unitary $U_j$. This results in $M_j=q_j^2 \one$,
hence the measurement probabilities are independent of $\rho$,
i.e. obtaining outcome $j$ reveals no information about the state of
the system. This ensures that $\rho_j=U_j\rho U_j^\dagger$.

In ADQC~\cite{Twisted2009,ADQC2010,Twisted2012}, the coupling and measurement
of a ancilla qubit to the system results in a two-outcome POVM where
the alternative Kraus operators are unitary and are related by a Pauli
correction. This requires the coupling between system and ancilla to
be of a special form, and that the ancilla qubit be prepared and
measured in particular directions~\cite{ADQC2010}. Should this not be
the case, then the effective Kraus operators may not result in the
desired unitary conditional evolution but may reveal information about
the system.

Without loss of generality, we may just consider two-outcome POVMs
since a multiple outcome POVM can be considered as the result of
multiple cascaded two-outcome POVMS~\cite{AO2008}. Using the singular
value decomposition, we can ignore the unitary transformations and
only consider the singular values which encode the (non)unitary
properties of the operator~\cite{DBN2013}.

The question we will answer is thus, given an initial POVM with Kraus
operators whose singular values are not equal, how can we perform
subsequent operations so that at least some of the outcomes result in
conditional unitary evolution of the initial state, and what is the
maximum probability of such corrective action?

We show that simple filtering or procrustean operations are sufficient
to ``equalize'' the cumulative singular values and thus unlearn the
information gained in prior steps. The maximum probability of enacting
conditional unitary evolution after an initial information-gaining
binary outcome measurement is related to the spectral width, i.e. the
difference between the largest and smaller singular values of the
Kraus operators.

\section{Procrustean Filtering}

We shall show how we can correct an initial non-unitary inducing
measurement by a filtering operation similar to that used for
entanglement concentration~\cite{Procrust1996}. Let us assume that
after the first measurement, we obtain the outcome associated with
Kraus operator $K_0=\text{diag}(q_0^k)$ where the $q_0^k$ are not all
the same. The probability of this result is $p_0=\tr[K_0^\dagger K_0
\rho]$ and varies from $(q_0^{k_{min}})^2\le p_0 \le
(q_0^{j_{max}})^2$. Since the probability depends on the state, we
gain information through this measurement and the state evolves
non-unitarily.

We now try to correct the evolution with
another measurement with diagonal Kraus operators
$K_{0,j}=\text{diag}(q_{0,j}^k),\;j=0,1$ resulting in the conditional
cumulative Kraus operators, $K_{0,j}K_0=\text{diag}(q_{0,j}^k
q_0^k)$. We can choose the singular values so that for one of the
outcomes, the resultant evolution is restored to being unitary.

Let us choose $K_{0,0}$ to correct $K_0$. If $q_0^{k_{min}}$ is the
smallest singular value of $K_0$, then setting
$q_{0,0}^k=q_0^{k_{min}}/q_0^k$ results in the cumulative operation
$K_{0,0}K_0=q_0^{k_{min}}\one$. The other outcome $K_{0,1}$ will have
at least one vanishing singular value, hence will have a non-trivial
nullspace and it will be impossible to further correct this branch of
the measurement tree. The probability of arriving at
$\tilde{K}_{0,0}=q_0^{k_{min}}\one$ is $p_{0,0}=(q_0^{k_{min}})^2$
independent of the initial state as required.

If at the first measurement we obtained the complementary result
$K_1=\text{diag}(q_1^k)$, then a subsequent correction would result in
outcome $\tilde{K}_{1,1}=q_1^{k_{min}}\one$ with probability
$p_{1,1}=(q_1^{k_{min}})^2$. The completeness of the measurement
operators implies that $(q_1^{k_{min}})^2=1-(q_0^{k_{max}})^2$, hence
the total probability of a successful correction after the initial
measurement is $p_{tot}=1-[(q_0^{k_{max}})^2-(q_0^{k_{min}})^2]$, or
one minus the visibility.

We can generalize the result to the case where the first measurement
has more than two outcomes. In this case, the probability of
successful correction is given by $p_{tot}=\sum_j
(q_j^{k_{min}})^2$. Hence the uncorrectable non-unitary action of the
initial measurement is determined by how much an outcome
\emph{excludes} a state compared with others.

\section{Partial Filtering}

\begin{figure}
\includegraphics[width=\columnwidth]{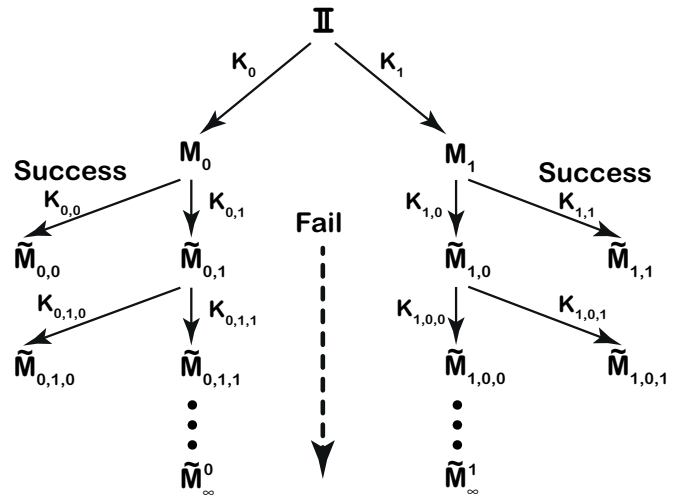}
\caption{Partial Filtering. Instead of succeeding or failing outright
  after one step, we can partially filter out corrected portions of
  the evolution, represented by the paths leading out to the
  sides. The vertical downward arrows represent partial failures, upon
  which we can retry recovery. The failure probability is given by the
  sum of the limiting residual cumulative measurement operators,
  $p_{fail}\one=\tilde{M}_{\infty}^0+\tilde{M}_{\infty}^1$.}
\label{fig:partial}
\end{figure}

We saw in the above section that we can choose our corrective
measurements to either succeed, or fail entirely with no further
recourse. An alternate strategy would be to succeed on one outcome,
but the alternative could still be further correctable. We shall
illustrate this in the case of a single qubit system.

Let the initial binary outcome measurement have Kraus operators,
$K_0=\text{diag}(a,b)$ and $K_1=\sqrt{\one-K_0^\dagger K_0}$ where
$0<b<a<1$. Suppose that we obtain outcome $K_0$, we can choose to
correct the evolution using the method in the previous section or else
we can choose, for example, the operators $K_{0,0}=\text{diag}(b,a)$
and $K_{0,1}=\sqrt{\one-K_{0,0}^\dagger K_{0,0}}$. In the case of the
result $K_{0,0}$, we achieve the cumulative evolution
$\tilde{K}_{0,0}=ab\one$, but the unsuccessful outcome
$\tilde{K}_{0,1}$ still has full rank and could be further
processed. The situation reduces to that of before but with a new
effective Kraus operator
$\tilde{K}_{0,1}=\text{diag}(a^{(1)}=a\sqrt{1-b^2},b^{(1)}=b\sqrt{1-a^2})$
and we can try to apply another round of corrections
(Fig.~\ref{fig:partial}).

This gives a recursive formula for the success probability for the
$K_0$ branch,
\begin{eqnarray}
p_{tot}^2=\sum_j p_j^0,\quad&&
p_j^0=\left(a^{(j)}b^{(j)}\right)^2\nonumber\\
a^{(j+1)}=a^{(j)}\sqrt{1-{b^{(j)}}^2},\quad&&
b^{(j+1)}=b^{(j)}\sqrt{1-{a^{(j)}}^2},
\end{eqnarray}
and for the $K_1$ branch,
\begin{eqnarray}
p_{tot}^1=\sum_j p_j^1,\quad&&
p_j^1=\left(c^{(j)}d^{(j)}\right)^2\nonumber\\
c^{(j+1)}=c^{(j)}\sqrt{1-{d^{(j)}}^2},\quad&&
d^{(j+1)}=d^{(j)}\sqrt{1-{c^{(j)}}^2},
\end{eqnarray}
where $c=\sqrt{1-a^2}$ and $d=\sqrt{1-b^2}$, and
$a^{(0)}=a$ \textit{etc}. It is simple to check that $a^{(j)}=d^{(j)}$ and
$b^{(j)}=c^{(j)}$ $\forall j\ge 1$.

In order to compute the limiting value of the success probability, it
is easier to compute the probability of failure. This can be found by
finding the limit of the unsuccessful Kraus operators given by
\begin{equation}
\tilde{K}_{\infty}^0=\text{diag}(a^{(\infty)},b^{(\infty)}),\quad
\tilde{K}_{\infty}^1=\text{diag}(c^{(\infty)},d^{(\infty)}),
\end{equation}
and the total failure probability is
\begin{equation}
p_{fail}\one=\tilde{M}_{\infty}^0+\tilde{M}_{\infty}^1
=\left({a^{(\infty)}}^2 + {b^{(\infty)}}^2\right)\one,
\end{equation}
where $\tilde{M}_{\infty}^{0,1}=(\tilde{K}_{\infty}^{0,1})^\dagger
\tilde{K}_{\infty}^{0,1}$

To solve the recursion formula, we first note that
${a^{(j+1)}}^2-{b^{(j+1)}}^2={a^{(j)}}^2-{b^{(j)}}^2=a^2-b^2$. We also
note that the fixed points of the recursion relation are when
$b^{(\infty)}=0$ leading to the limit
\begin{equation}
K_{\infty}^0=\text{diag}(a^2-b^2,0),\quad
K_{\infty}^1=\text{diag}(0,a^2-b^2),
\end{equation}
hence $p_{fail}=a^2-b^2$, conversely $p_{tot}=1-(a^2-b^2)$, the same
as for the Procrustean method.

\section{Success Bound}
\label{sec:success}

We show that the Procrustean method achieves the maximum probability
of success. In general, assume that at some stage of the measurement
tree the effective Kraus operator is given by
$K_j=\text{diag}(q_j^k)$. Any corrective set of Kraus operators has to
satisfy $\one=\sum_k K_{j,k}^\dagger K_{j,k}$ which implies that
\begin{equation}
\sum_k \tilde{K}_{j,k}^\dagger \tilde{K}_{j,k}=K_j^\dagger K_j,
\end{equation}
where $\tilde{K}_{j,k}=K_{j,k}K_j$ is the cumulative Kraus operator
for the outcome $k$.

If we consider all branches $k'$ that result in conditional unitary
evolution, then these add up to
$\sum_{k\in\{k'\}}\tilde{K}_{j,k}^\dagger
\tilde{K}_{j,k}=p_{j}^{succ}\one$, hence the branches that do not
succeed sum to $K_j^\dagger K_j-p_{j}^{succ}\one$. As this has to be a
positive operator, $p_{j}^{succ}$ cannot be larger than the square of
the minimum singular value of $K_j$. The procrustean method saturates
this bound. We note that this result is independent of the use of the
singular value decomposition (one may work just with the measurement
operators) hence encompasses any general set of correction Kraus
operators, not just those ``aligned'' with the bases of previous
results.

When considering all of the initial branches of an non-ideal
measurement $\{M_j\}$, the maximum total probability of recovery is
given by $\sum_j p_{j}^{min}$ where $p_j^{min}$ is the minimum
probability to obtain outcome $j$ when taken over all possible input
states. Measurement operators of the form
$M_j=q_j^2(\one-\proj{\psi_j})$ reveal little information, especially as
the dimensionality of the space increases, but are not recoverable.

\section{Application to probabilistic teleportation}

We apply the results to the well studied problem of probabilistic
quantum teleportation as an illustration. Alice and Bob share a
non-maximally entangled state of the form
$\ket{\Psi(\theta)}=\cos\frac{\theta}{2}\ket{00}+\sin\frac{\theta}{2}\ket{11}$
where $0\le\theta\le\pi/2$. Charlie gives Alice a qubit in the state
$\ket{\phi}=\alpha\ket{0}+\beta\ket{1}$ to teleport to Bob with the proviso that
it either arrives with unit fidelity, or else it fails. The standard
solution~\cite{AP2002} is for Alice to measure in a non-maximally entangled basis,
\begin{eqnarray}
\ket{\Psi^{(0)}}&=&\sin\frac{\theta}{2}\ket{00}+\cos\frac{\theta}{2}\ket{11}\nonumber\\
\ket{\Psi^{(1)}}&=&\sin\frac{\theta}{2}\ket{10}+\cos\frac{\theta}{2}\ket{01}\nonumber\\
\ket{\Psi^{(2)}}&=&\cos\frac{\theta}{2}\ket{00}-\sin\frac{\theta}{2}\ket{00}\nonumber\\
\ket{\Psi^{(3)}}&=&\cos\frac{\theta}{2}\ket{10}-\sin\frac{\theta}{2}\ket{01}.
\end{eqnarray}
The first two outcomes will be obtained each with probability
$\frac{1}{4}\sin^2 \theta$ and result in (reversible) unitary quantum
channels between Alice and Bob.

For the other two results, Alice obtains some information about
Charlie's state resulting in operations with singular values
$\{\cos^2\frac{\theta}{2},\sin^2\frac{\theta}{2}\}$. Bob can choose to
reverse the non-unitary dynamics by filtering with probability $\sin^4
\frac{\theta}{2}$ in both cases. The total probability of Alice and
Bob to succeed in teleporting $\ket{\phi}$ is
$p=2\left(\frac{1}{4}\sin^2 \theta\right)+2\left(\sin^2
  \frac{\theta}{2}\right)^2=1-\cos\theta$. This recovery is optimal as
it matches the sums of the squares of the minimal singular values of
the initial $4$-outcome POVM on $\ket{\phi}$. We note that this
probability matches that of initially filtering $\ket{\Psi(\theta)}$
to obtain a maximally entangled state prior to conventional
teleportation.

\section{Conclusion and Discussion}

These results answer a question about general binary measurement trees
and the form that they can take~\cite{AO2008}. The extension of the
success bound to arbitrary measurement trees implies that trees with
all the final operators conditionally unitary cannot have any
non-unitary branch within it. This places strong constraints on the
allowed couplings in ADQC-like architectures as all ancilla-driven
dynamics much be unitary to maintain the continuing coherence of the
register~\cite{Oi2014}. Even relaxing the requirement for
determinism~\cite{SO2013A}, the Cartan decomposition of the
system-ancilla interaction must remain rank deficient, i.e. not of the
SWAP form~\cite{Cartan2003}.

The maximum probability of recovery takes on a simple form when
restricted to a binary outcome POVM, being the one minus the
difference between the maximum and minimum measurement
probabilities. In the case of a multiple outcome POVM, it becomes the
sum of the minimum probabilities of each measurement operator.

The results presented recreate those given in Refs.~\cite{KU1999}
and~\cite{CL2012} but we make minimal reference to states, the
emphasis here is entirely on the Kraus operators and elements of the
effective POVM. In this way, the proof presented in
Sec.~\ref{sec:success} is considerably shortened and simplified
compared with previous papers. Also in contrast, instead of
information gain in terms of estimation fidelity, the results suggest
that reversibility is better characterized by the ability to discount
the possibility of a state or subspace~\cite{BJOP2014}. For example in
the continuous variable case, an overcomplete POVM with uncountably
many elements all that are proportional to $\one-\proj{\alpha}$
($\ket{\alpha}$ is a coherent state)~\cite{OPJ2013}, would reveal
little information about an input state, but would not be reversible
for any of its outcomes.

\begin{acknowledgments}
  DKLO acknowledges fruitful discussion with John Jeffers, and is
  supported by Quantum Information Scotland (QUISCO).
\end{acknowledgments}


\begin{thebibliography}{99}

\bibitem{MBQC2003}
R. Raussendorf, D. E. Browne and H. J. Briegel, Phys. Rev. A
  \textbf{68}, 022312 (2003)

\bibitem{Twisted2009}
E. Kashefi, \textit{et al.}, 
Electronic Notes in Theoretical Computer Science \textbf{249}, 307-331 (2009)

\bibitem{ADQC2010}
  J. Anders, \textit{et al.},
  Physical Review A \textbf{82}, 020301 (2010)

\bibitem{Twisted2012}
  J. Anders, \textit{et al.}, 
Th. Comp. Sci. \textbf{430}, 51 (2012)

\bibitem{AP1989}
J. Anandan and A. Pines, Phys. Lett. A \textbf{141},
  335 (1989)

\bibitem{Oi2014}
D. K. L. Oi, Phys. Rev. A, In Press (2014), arxiv:1402.1104

\bibitem{MK2010}
T. Morimae, J. Kahn,
Phys. Rev. A \textbf{82}, 052314 (2010)

\bibitem{RHG2007}
R. Raussendorf, J. Harrington and K. Goyal,
New J. Phys. \textbf{9}, 199 (2007)


\bibitem{UIN1996}
M. Ueda, N. Imoto, and H. Nagaoka, Phys. Rev. A \textbf{53}, 3808 (1996)

\bibitem{KU1999}
M. Koashi and M. Ueda,
  Phys. Rev. Lett. \textbf{82}, 2598 (1999)

\bibitem{ParaoanuPRA2011}
G. S. Paraoanu, Phys. Rev. A \textbf{83}, 044101 (2011)


\bibitem{CL2012}
  Y. W. Cheong and S.-W. Lee, Phys. Rev. Lett. \textbf{109},
  150402 (2012)


\bibitem{KJ2006}
A. N. Korotkov, A. N. Jordan, Phys. Rev. Lett. \textbf{97}, 166805 (2006)

\bibitem{JK2010}
A. N. Jordan, A. N. Korotkov, Contemporary Physics \textbf{51}, 125 (2010)

\bibitem{ParaoanuEPL2011}
  G. S. Paraoanu, Euro. Phys. Lett \textbf{93}, 64002 (2011)


\bibitem{Katz2008}
Nadav Katz \textit{et al.}, 
Phys. Rev. Lett. \textbf{101}, 200401 (2008)

\bibitem{Kim2009}
Yong-Su Kim, Young-Wook Cho, Young-Sik Ra, and Yoon-Ho Kim,
Optics Express \textbf{17}, 11978 (2009)



\bibitem{AO2008}
E. Andersson and D. K. L Oi, Phys. Rev. A \textbf{77}, 052104 (2008)


\bibitem{DBN2013}
J. Dressel, T. A. Brun and A. N. Korotkov, arXiv:1312.1319

\bibitem{Procrust1996}
C. H. Bennett, H. J. Bernstein, S. Popescu and B. Schumacher,
Phys. Rev. A \textbf{53}, 2046 (1996)

\bibitem{AP2002}
P. Agrawal and A. K. Pati,
Phys. Lett. A \textbf{305}, 12 (2002)


\bibitem{SO2013A}
K. Halil-Shah, D. K. L. Oi, Proc. TQC 2013, LIPIcs \textbf{22}, 1 (2013)

\bibitem{Cartan2003}
J. Zhang, J. Vala, S. Sastry and K. B. Whaley,
Phys. Rev. A \textbf{67}, 042313 (2003)





\bibitem{BJOP2014}
S. Bandyopadhyay, R. Jain, J. Oppenheim and C. Perry,
Phys. Rev. A \textbf{89}, 022336 (2014)

\bibitem{OPJ2013}
D. K. L. Oi, V. Potocek and J. Jeffers,
Phys. Rev. Lett. \textbf{110}, 2010504 (2013)

\end{thebibliography}
\end{document}